\documentclass[%
 aip,
 amsmath,amssymb,
 reprint,%
]{revtex4-1}

\usepackage{graphicx}
\usepackage{dcolumn}
\usepackage{bm}

\usepackage[utf8]{inputenc}
\usepackage[T1]{fontenc}
\usepackage{mathptmx}
\usepackage{etoolbox}

\makeatletter
\def\@email#1#2{%
 \endgroup
 \patchcmd{\titleblock@produce}
  {\frontmatter@RRAPformat}
  {\frontmatter@RRAPformat{\produce@RRAP{*#1\href{mailto:#2}{#2}}}\frontmatter@RRAPformat}
  {}{}
}%
\makeatother
\begin{document}

\preprint{AIP/123-QED}

\title[UEDS as a Tool for Studying Phonon Transport]{Ultrafast Electron Diffuse Scattering as a Tool for Studying Phonon Transport: Phonon Hydrodynamics and Second Sound Oscillations}

\author{Laurenz Kremeyer}
\affiliation{Department of Physics, Center for the Physics of Materials, McGill University, H3A\,2T8, Montreal, Canada.}

\author{Tristan L. Britt}
\affiliation{Department of Physics, Center for the Physics of Materials, McGill University, H3A\,2T8, Montreal, Canada.}

\author{Bradley J. Siwick}
\affiliation{Department of Physics, Center for the Physics of Materials, McGill University, H3A\,2T8, Montreal, Canada.}
\affiliation{Department of Chemistry, McGill University, H3A\,0B8, Montreal, Canada.}

\author{Samuel C. Huberman}
\affiliation{Department of Chemical Engineering, McGill University, H3A\,0C5, Montreal, Canada.}
\email{samuel.huberman@mcgill.ca}

\date{\today}

\begin{abstract}
Hydrodynamic phonon transport phenomena, like second sound, have been observed in liquid Helium more than 50 years ago.
More recently second sound has been observed in graphite at over 200\,K using transient thermal grating (TG) techniques.
In this work we explore signatures of phonon hydrodynamic transport and second sound oscillations in ultrafast electron diffuse scattering (UEDS) patterns, which can provide time, momentum and branch resolved information on the state-of-excitation of the phonon system beyond that available through TG experiments.
We use density functional theory and solve the Boltzmann transport equation to determine time-resolved non-equilibrium phonon populations and model phonon transport in graphite.
This model also provides the information necessary to calculate the time evolution of one-phonon structure factors and diffuse scattering patterns during thermal transport covering ballistic, diffusive, and hydrodynamic regimes where the effect of a second sound oscillation on the phonon distribution is observed.
Direct measurements of how the phonon distribution varies in time and space in various thermal transport regimes should yield new insights into the fundamental physics of the underlying processes.
\end{abstract}

\maketitle

\section{Introduction}
The conventional approach to modeling heat transport in a solid relies on Fourier's law
\begin{equation}
Q = -k \cdot \nabla T,
\end{equation}
where the local heat flux $Q$ and the temperature gradient $\nabla T$ are directly proportional, related to one another by the thermal conductivity $k$.
This phenomenological equation provides a reasonable description of thermal transport in the diffusive regime, but breaks down when the length or time scales approach those of the microscopic phonon scattering processes. 
There are, in fact, at least three distinct thermal transport regimes\cite{prohofsky1964,cepellotti2015}.
In the diffusive regime, non-momentum-conserving scattering events such as Umklapp and isotope scattering are dominant.
In the ballistic regime, extrinsic scattering events at surfaces, boundaries or defects are dominant over all other scattering mechanisms\cite{lane1947}.
In between the diffusive and ballistic regime, may lie the hydrodynamic regime, contingent on the material, where momentum-conserving normal scattering events are dominant compared to Umklapp- and isotope-scattering.
This momentum-conservation supports the existence of hydrodynamic behavior and analogies to classical fluids emerge.
One of the observable signatures of the phonon hydrodynamic regime is a phenomenon called \emph{second sound}\cite{peshkov1944}.
Second sound is associated with the wave-like motion of temperature, which has a lower velocity than the \emph{first} speed of sound of the material and thus lags behind propagating elastic waves (hence the term \emph{second} sound).

Recent experiments have observed hydrodynamic behavior in graphite above 100\,K\cite{huberman2019} and 200\,K\cite{ding2022}.
Furthermore, some 2D materials such as graphene have been theoretically shown support the same phenomena\cite{cepellotti2015}.
These recent developments suggest a need for experimental techniques that can better resolve the microscopic phonon dynamics that occur in the phonon hydrodynamic regime.
One such technique is ultrafast electron diffuse scattering (UEDS) which provides time- and momentum-resolved information on the nonequilibrium state of the phonon system in materials.
In certain cases it is even possible to disentangle the contributions to the scattering pattern by phonon-branch.
This technique has previously provided momentum-resolved information on inelastic electron-phonon coupling\cite{durr2021,rdc2022,stern2018,seiler2021}, demagnetization dynamics\cite{durr2020}, soft phonon and charge density wave physics\cite{otto2021,cheng2022} in bulk materials and monolayers\cite{britt2022}.
Since heat in non-metallic materials in mainly transported through the phonons, UEDS is also capable to investigate exotic heat transport regimes by interrogating the nonequilibrium state of the phonon system directly.

\begin{figure}[b]
\includegraphics[width=\columnwidth]{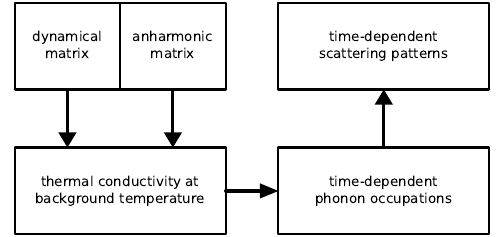}
\caption{Flowchart of the computational procedure to obtain scattering patterns.}
\label{fig:flowchart}
\end{figure}

Here we demonstrate the potential for UEDS in studying a variety of phonon transport regimes, but most notably the phonon hydrodynamic regime.
We present a novel approach to compute the time-resolved, phonon-diffuse scattering intensities in response to the phonon system being driven out of equilibrium by laser-heating.
The full scattering matrix Boltzmann transport equations are solved, and subsequently the one-phonon structure factors are calculated.
The output of these computations can be directly compared with electron/x-ray pump probe experiments.
We restrict our focus to the transient thermal grating (TG) geometry in graphite.
Graphite has previously been studied by UEDS\cite{stern2018, rdc2019} and via TG where second sound oscillations were observed\cite{huberman2019,ding2022}.
This approach, however, is general and can be extended to study other heating geometries (e.g.~a Gaussian heat source) and other materials (e.g.~graphene and MoS$_2$).  

\section{Methods}
We combine the Green's function approach to solve the linearized Boltzmann transport equation (BTE) with one-phonon structure factor calculations to predict the observables of a pump-probe UEDS dataset to demonstrate the ability of UEDS to resolve heat transport at the microscopic level.
A high-level overview of the computational procedure is provided in Figure~\ref{fig:flowchart}.

\subsection{Solving the linearized BTE}
\begin{figure*}[t]
\includegraphics[width=\textwidth]{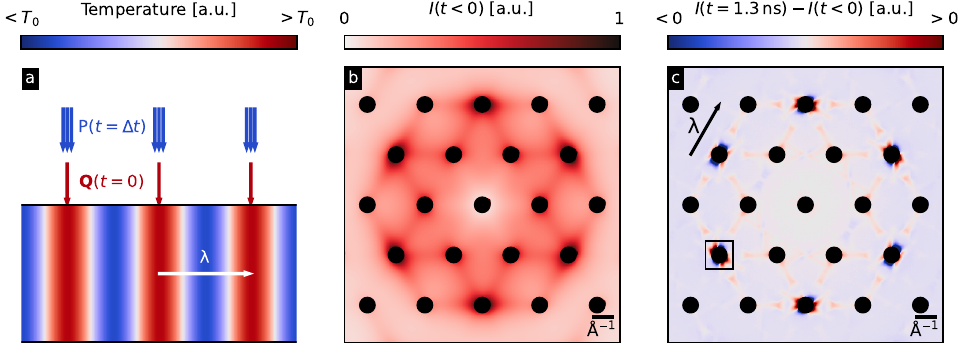}
\caption{(a)~Schematic setup of the simulated pump-probe diffraction experiment in real space. The sample in the lower half of the panel is exposed to the heat flux $\mathbf{Q}$ at $t=0$. Here we choose a sinusoidal temperature grating across the system with a wave vector $\lambda$ as has been generated by laser excitation in previous all-optical experiments\cite{ding2022}. Blue areas are colder than the background temperature $T_0$, red ones are hotter. At a later point in time $t=\Delta t$ the initially hot parts of the sample are probed with electron pulses $P$. (b)~Simulated diffuse scattering pattern of graphite in equilibrium at 100\,K. (c)~Differential diffuse scattering intensity $I(t<0) - I(t=1.3\,\mathrm{ns})$ after imposing a sinusoidal heat profile with a grating period of $\lambda=$10\,\textmu m onto the sample. The time chosen (1.3 ns) is approximately the first minimum associated with the second sound oscillation shown more clearly in Figures~\ref{fig:diff_vs_time} and \ref{fig:correlations}. Black dots indicate the regions near $\Gamma$ points where the bright Bragg peaks appear. The arrow denotes the direction of the temperature grating.}
\label{fig:eq_vs_diff}
\end{figure*}

The commonly used relaxation time approximation (RTA) to the BTE breaks down for low temperatures or materials with high Debye temperatures like graphite and graphene.
To capture and predict phonon hydrodynamics, one must solve, at a minimum the linearized BTE with a full scattering matrix version of the collision operator, which is until recently, has been a formidable challenge.
The Green's function solution of the linearized BTE has recently been obtained, enabling the accurate prediction of not only temperature, but also individual phonon distribution functions, in response to arbitrary spatiotemporal heating profiles \cite{chiloyan2021}.
We apply this approach here for the case of a temporally impulsive excitation, mimicking the excitation event of a pump-probe experiment that has a sinusoidal spatial profile as shown in Figure~\ref{fig:eq_vs_diff}\,(a).
Beginning with the linearized BTE 
\begin{equation}
\label{eq:lbte}
\frac{\partial g_n}{\partial t} + \mathbf{v}_n \cdot \vec{\nabla} g_n = Q_n + \sum_j \omega_n \mathbf{W}_{n, j} \frac{1}{\omega_j} \left( c_j \Delta T - g_j, \right)
\end{equation}
and following the solving procedure described in Ref.\cite{chiloyan2021}, yields time-dependent deviational phonon energy densities per mode~$n$, defined as
\begin{equation}
\label{eq:gn}
g_n \equiv \frac{\hbar \omega_n}{N V} \left(f_n - \frac{1}{\mathrm{e}^{\frac{\hbar \omega_n}{k_\mathrm{B}T_0}} - 1}\right).
\end{equation}
$\{\mathbf{v}_n\}$ are the phonon group velocities, $\{Q_n\}$ are the volumetric heat generation rates, $\{\omega_n\}$ are the phonon mode frequencies, $\mathbf{W}_{n, j}$ is the phonon phonon scattering matrix, $\{c_j\}$ are the phonon heat capacities and $\Delta T$ is the temperature difference with respect to the background temperature $T_0$.
The sum runs over all phonon branches $j$.
$N$ is the number of discretized points in the Brillouin zone, $V$ is the unit cell volume and $\{f_n\}$ are the phonon occupations.

Assuming that the phonon distributions only deviate slightly from their equilibrium, we can linearize the Bose-Einstein distributions in terms of temperature variation as
\begin{equation}
f^0_n \approx \frac{1}{\mathrm{e}^{\frac{\hbar \omega_n}{k_\mathrm{B}T_0}} - 1} + \frac{\hbar \omega_n}{N\,v}c_n \Delta T,
\end{equation}
with the temperature change $\Delta T$.
When combined with the assumption that $\frac{\Delta T}{T_0}$ is small, the elements of the full scattering matrix depend only on the background temperature and not the temperature change. 

To obtain the inputs Eq.~\ref{eq:lbte}, the second and third order force constants are first computed with the \texttt{PHonon} and \texttt{D3Q} packages of the density functional theory suite Quantum Espresso\cite{QE1, QE2, D3Q} along with \texttt{thirdorder.py}\cite{li2014shengbte}. Details of the calculations can be found in Ref.~\cite{huberman2019}.

\subsection{Computing electron/x-ray diffuse scattering patterns}
To predict experimental diffuse scattering data from the results of the BTE calculations we can, under the kinematical approximation, write the scattered intensity as
\begin{equation}
I(\mathbf{q}, t) = I_0(\mathbf{q}, t) + I_1(\mathbf{q}, t) + \hdots,
\end{equation}
with the intensity contribution of scattering events involving $\mathrm{j}$ phonons $I_\mathrm{j}$ at time $t$ at the scattering vector $\mathbf{q}$.
The zero-phonon term
\begin{equation}
I_0(\mathbf{q}, t) \propto \sum_m \sum_s \sum_{s\prime} \left[ f_s f_{s\prime} \mathrm{e}^{-M_s-M_{s\prime}\mathrm{e}^{-\mathrm{i}\mathbf{q}\cdot(\mathbf{R}_m + \mathbf{\boldsymbol\tau}_{s,s\prime})}} \right],
\end{equation}
with the Debye-Waller factors
\begin{equation}
M_s = \frac{1}{4\mu_s} \sum_\mathbf{k} \sum_j \left| a_{\mathbf{k}, j} \right|^2 \left| \mathbf{q} \cdot \mathbf{e}_{\mathbf{k}, j, s} \right|^2
\end{equation}
and the unit cells $\{m\}$ and the atoms $\{s\}$, the atomic structure factors $\{f_s\}$, the origins of the lattice vectors of the unit cells $\{\mathbf{R}_m\}$, the atomic basis vectors $\{\boldsymbol\tau_s\}$, the atomic masses $\{\mu_s\}$ the wave vectors in the first Brillouin zone $\{\mathbf{k}\}$, the phonon branches $\{j\}$ the vibration amplitudes $\{a_{\mathbf{k}, j}\}$ and the polarization vectors $\{\mathbf{e}_{\mathbf{k},j,s}\}$ corresponds to Bragg scattering\cite{xu2005}.
The one-phonon term captures the majority of the diffuse intensity that is scattered into the regions between the Bragg peaks.
Higher order terms can play a significant role in diffuse scattering\cite{zacharias2021_a, zacharias2021_b}, but they are neglected due to the low temperatures used in the simulations.
The one-phonon scattering intensity has also been derived by Xu and Chiang\cite{xu2005} and by plugging in equation~\ref{eq:gn} works out to be
\begin{equation}
I_1(\mathbf{q}, t) \propto \sum_n \left( g_n(\mathbf{q}, t) \frac{NV}{\hbar \omega_n(\mathbf{q})} + \frac{1}{\mathrm{e}^{\frac{\hbar \omega_n(\mathbf{q})}{k_\mathrm{B}T_0}} - 1} + \frac{1}{2}\right) \frac{\left| F_{1, n}(\mathbf{q}) \right|^2}{\omega_n(\mathbf{q})},
\end{equation}
where $F_{1, n}(\mathbf{q})$ is the one-phonon structure factor that weighs the contribution of scattering from each phonon mode at different scattering vectors with the phonon polarization vector in a scalar product $(\mathbf{q}\cdot\mathbf{e}_n)$.
The zero-phonon intensity contribution is dependent on the collective motion of all atoms in the sample and their mean square displacement (i.e., the Debye-Waller effect).
Since this is directly related to the temperature of the sample, one might use the transient intensity of the Bragg peaks to track temperature changes in the sample and compare the results to experimental data \cite{huberman2019,ding2022}.
The one-phonon intensity, however, allows for a more sophisticated analysis of the phonon system, because of the $(\mathbf{q}\cdot\mathbf{e}_n)$ selection rule, enabling access to mode-level information.
Based on this insight, for simple crystal systems, a linear system of equations can be constructed that allows for every single phonon mode population to be tracked in time\cite{rdc2019}.

Since the simulated diffraction patterns are selectively computed for the initially hot parts of the system, it is important to also design an experimental setup which allows to only scatter electrons/X-rays off of equivalent sections of the temperature grating.
One could use a strongly focused electron/X-ray beam or employ a masking approach.
The heating profile could be created with interference techniques similar to the one used in \cite{huberman2019}.
In the computations, we use the source terms $\{Q_\mathrm{n}\}$ in Equation~\ref{eq:lbte} to create a temperature grating that is homogeneous in out-of-plane directions.
We justify this assumption by the fact that very thin samples have been used in UEDS/X-ray scattering experiments\cite{stern2018, britt2022}, and in this case, the low out-of-plane thermal conductivity of graphite\cite{zhao2020} does not pose a problem.

\begin{figure*}[t]
\includegraphics[width=\textwidth]{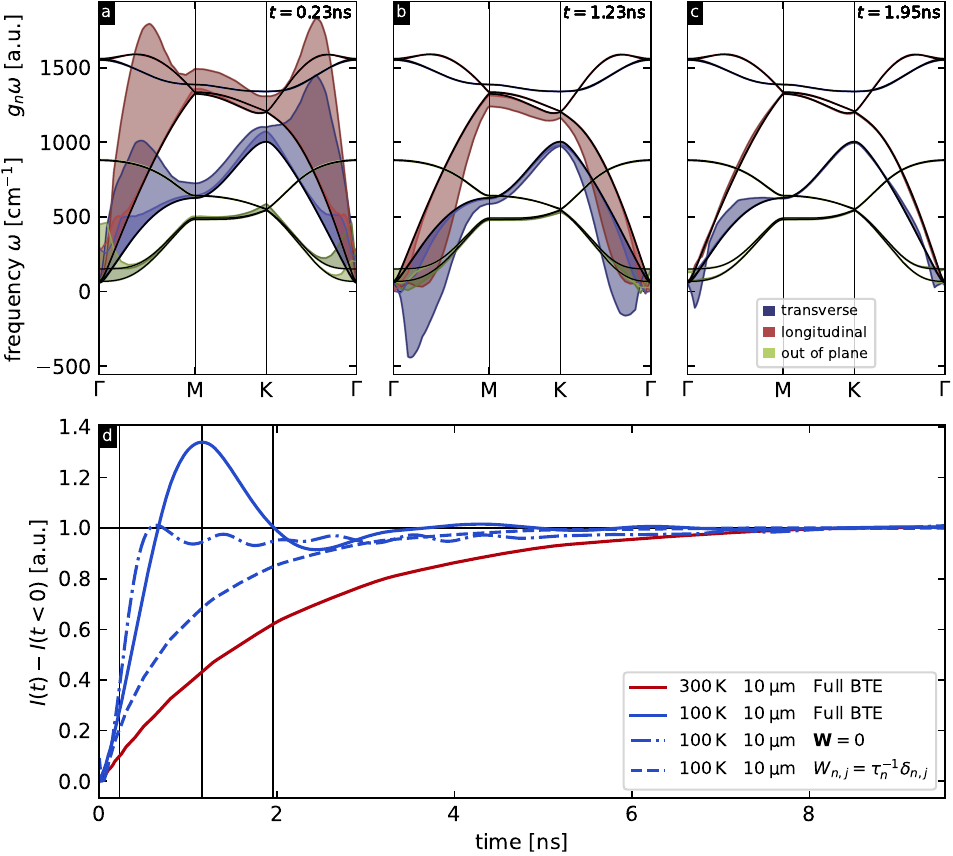}
\caption{(a)-(c)~Phonon band structure of graphite along $\overline{\Gamma \mathrm{M} \mathrm{K} \Gamma}$ shown with black solid lines. The shaded areas indicate the (nonequilibrium) deviation in the energy stored $g_n\omega$ in the LA~(red), TA~(blue) and ZA~(green) phonon modes compared to equilibrium at 100\,K for the three time steps marked by vertical black lines in panel~(d). Shaded areas above/below the black lines indicate positive/negative deviations in excitation energy. Negative deviations are a clear violation of Fourier's law (heat diffusion) and are an indication of hydrodynamic/oscillatory transport (d)~Integrated Bragg peak intensity the (220) peak (indicated with the black box in Figure~\ref{fig:eq_vs_diff}\,(b)) versus time for four different simulation conditions. Solid lines depict the results for the solutions with the full scattering matrix, for the dash-dotted line the scattering matrix is set to zero and the dashed line represents the RTA solution. A second sound oscillation is evident at 100\,K, indicating hydrodynamic transport, but not at 300\,K where diffusive transport results in a monotonic decay. The transition between these two regimes occurs naturally in the BTE framework as the rate of Umklapp scattering in graphite increases with background temperature.}
\label{fig:diff_vs_time}
\end{figure*}

\section{Results}
Based upon the above approach, we explore the signatures of the three microscopic heat transport regimes described in single crystal graphite when studied via UEDS.
The electron beam is taken to be incident along the [001] zone-axis, or perpendicular to the graphene sheets (i.e. along the c-axis of the graphite crystal).
In this geometry, phonon modes polarized out-of-plane are not observable and are not considered further.
Figure~\ref{fig:eq_vs_diff}\,(b) shows an equilibrium diffuse scattering pattern of graphite at 100\,K.
The diffuse intensity is four to six orders of magnitude weaker than the Bragg peak intensity.
Panel~(c) depicts how the diffuse scattering intensity changes 1.3\,ns after exciting it with the sinusoidal heating profile with a grating period of $\lambda$=10\,\textmu m. Blue areas lose intensity, whereas red areas gain intensity.
In the shown calculations, the energy deposited into the system via the source terms $\{Q_n\}$ (laser excitation) is distributed between all modes across whole Brillouin zone according to their respective heat capacities.
This is a justifiable approximation of the state of the phonon system $> 200$\,ps following photoexcitation\cite{rdc2019}.
Other initial energy distributions across the Brillouin zones have been simulated (e.g., modes near the K-points strongly excited), but phonon-phonon scattering is found to be much faster than the relevant timescales for the transport properties of interest and the results shown here are not particularly sensitive to the initial conditions taken for the state of the phonon system.
The effects are most pronounced around the second order Bragg peaks close to the $\Gamma$ points, which hints at a major contribution to the thermal transport from the acoustic modes near the zone center.
In addition, the six-fold symmetry observed in the equilibrium pattern that reflects the symmetry of the graphite crystal is broken and the differential pattern only shows a two-fold symmetry along the temperature grating wave vector $\lambda$.

The evolution of the deviational phonon energy densities obtained as solutions to the linearized BTE weighted by their respective phonon frequencies is depicted in Figure~\ref{fig:diff_vs_time}\,(a)-(c).
The background temperature is 100\,K and the spatial periodicity of the temperature grating is 10\,\textmu m.
Panel~(a) represents the state of the system shortly after excitation (0.23\,ns), panel (b) at the minimum sample temperature (1.23\,ns) and panel~(c) at the second time the sample temperature reaches the background temperature (1.95\,ns).
At the earliest time, many phonon modes are highly populated across the Brillouin zone.
From comparing panels~(a) and~(b) it is clear the modes depopulate and phonon-phonon scattering leads to a general relaxation towards the zone-center.
At larger times, the transverse acoustic mode is the dominant contributor to hydrodynamic heat transport.
The higher frequency optical modes have very low deviational phonon energy densities and their contribution to the heat transport is thus negligible.

The transient behavior of the Bragg peaks is depicted in Figure~\ref{fig:diff_vs_time}\,(d).
The plot shows the integrated differential transient intensity of a second order Bragg peak for different simulation conditions.
Using the method as described above we set the background temperature of the system to 300\,K and use a 10\,\textmu m grating period.
The observed behavior shows an approximate exponential decay towards the equilibrium intensity.
Hence the system is cooling down from the excited state to the background temperature which is expected in the diffusive regime where the rate of Umklapp scattering is comparable or greater than that of Normal scattering.
Decreasing the background temperature to 100\,K depopulates higher momentum phonon modes and thus reduces the Umklapp scattering rate.
The transient intensity shows oscillations that are indicative of hydrodynamic heat transport.
From the oscillation period and the grating size we can calculate the speed of second sound to be $\frac{5\text{\textmu m}}{1.23\,\mathrm{ns}}=4065\,\frac{\mathrm{m}}{\mathrm{s}}$, which is about a factor of 3.5 to 5.5 times slower than the speed of sound of the transverse and longitudinal phonon branches in graphite\cite{bosak2007}.
The oscillations in our data are equivalent to what was experimentally observed in transient TG experiments\cite{huberman2019}.

\begin{table}[t]
\caption{Dataset parameters and corresponding heat transport regimes.}
\label{tab:regimes}
\begin{ruledtabular}
\begin{tabular}{rll}
$T_0$ & Method & Regime \\
\hline \\[-3mm]
300 & Full BTE & Diffusive \\
100 & Full BTE & Hydrodynamic\\
100 & $\mathbf{W}=0$ & Ballistic \\
100 & $W_{n, j} = \tau_n^{-1} \delta_{n,j}$ & Nonphysical\\
\end{tabular}
\end{ruledtabular}
\end{table}

Further we show that the hydrodynamic case cannot be accurately simulated using the relaxation-time approximation in which the scattering matrix is only non-zero for the diagonal entries and thus there is no inter mode scattering and each mode has its own independent lifetime, which is nonphysical as it leads to energy conversation violation.
Clearly, second sound is a collective, many-body phenomenon that emerges through momentum-conserving interactions between phonons described in detail by the scattering matrix~$\mathbf{W}$. 
To produce a dataset that covers the ballistic transport regime the scattering-matrix is set to zero.
This leads to a faster temperature decay than in the other cases; note that the intensity shows small amplitude high frequency oscillations that decay slowly, which can partly be traced back to numerical artifacts.
We note that the analytical solution to ballistic transport in the transient grating geometry is a cosine integral over the Brillouin zone with the characteristic frequencies determined by a dot product between the group velocity and grating wavevector, which is a slowly converging function.
Ding et al.\cite{ding2022} recently predicted a similar oscillatory behavior in the ballistic regime in graphite.
Theoretical predictions demonstrate that whether or not oscillatory behavior is observed in the ballistic regime depends on the phonon dispersion and temperature\cite{zhang2022emergence}, which is in contrast to the hydrodynamic regime where oscillatory behavior is always observed.
Disentangling the contributions requires further experimental work.
A summary of the calculation parameters and their attributed heat transport regime is provided in Table~\ref{tab:regimes}.
Snapshots of the deviational phonon populations analogous to panels~(a)-(c) for the diffusive and ballistic cases can be found in the supplementary material\cite{SM}.

\begin{figure*}[t]
\includegraphics[width=\textwidth]{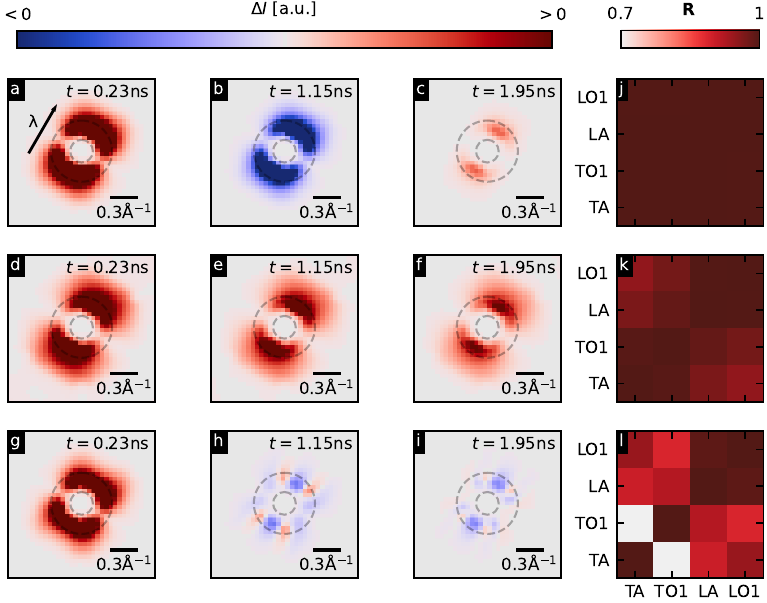}
\caption{(a)-(i)\,Scattering contribution from the transverse acoustic~(TA) mode in the section of the diffraction pattern depicted with the black box in Figure~\ref{fig:eq_vs_diff} for the hydrodynamic case ((a)-(c)), the diffusive case ((d)-(f)) and the ballistic case ((g)-(i)) for the timesteps $\{0.22\,\mathrm{ns}, 1.23\,\mathrm{ns}, 1.95\,\mathrm{ns}\}$ respectively. The arrow in (a) depicts the direction of the wavevector of the temperature grating. (j)-(l)~Pearson correlation coefficient matrix for the integrated intensity in a ring from 0.13\,\AA$^{-1}$ to 0.35\,\AA$^{-1}$ between TA, first transverse optical~(TO1), longitudinal acoustic~(LO) and first longitudinal optical mode~(LO1) in the first 1.75\,ns after excitation for the hydrodynamic case (j), the solution using the RTA (k) and the ballistic case (l). The ring is shown with opaque dashed lines in (a)-(i).}
\label{fig:correlations}
\end{figure*}

We shall now shift our focus from the Bragg peaks to the diffuse scattering to obtain phonon momentum-resolved insights into the transport phenomena than could otherwise not be obtained by simply making measurements that are proportional to the local sample temperature or performing conventional TG measurements. 
Figure~\ref{fig:eq_vs_diff}\,(c) shows that the diffuse scattering intensity is most pronounced near the $\Gamma$ points, so in the following we will exclusively discuss these regions.
Figure~\ref{fig:correlations}\,(a)-(i) shows the contribution to the diffuse scattering pattern close to the Gamma point in the area that is marked by the black box in Figure~\ref{fig:eq_vs_diff}\,(c) by the transverse acoustic phonon branch.
In the diffusive ((d)-(f)) case the diffuse intensity rises across the whole Brillouin zone and the phonon populations do not drop below the equilibrium value at any point in time, whereas there are increases and decreases in intensity for the hydrodynamic ((a)-(c)) and ballistic ((g)-(i)) cases.
It is striking that the ballistic regime shows a much more complex structure in the scattering intensity around $\Gamma$.
An animated video for all the simulated timesteps can be found in the supplementary material\cite{SM}.

Correlating the scattering intensity contributions of the four lowest energy phonon branches gives insight into the momentum-conserving heat transport in system.
Specifically, we compare the transient intensities $\{I^\odot_n\}$ given by
\begin{equation}
I^\odot_n = \int_\odot I_n(\mathbf{q}, t) \mathrm{d}\mathbf{q}.
\end{equation}
The integral runs over to domain $\odot$ which corresponds to the rings depicted in Figure~\ref{fig:correlations}\,(a)-(c).
The matrices in (d-f) show the Pearson correlation coefficients
\begin{equation}
R_{n,j} = \frac{\sum\limits_{t=0}^{1.75\,\text{ns}}(I^\odot_n - \overline{I^\odot_n})(I^\odot_j - \overline{I^\odot_j})}{\sqrt{\sum\limits_{t=0}^{1.75\,\text{ns}}(I^\odot_n - \overline{I^\odot_n})^2 \quad  \sum\limits_{t=0}^{1.75\,\text{ns}}(I^\odot_j - \overline{I^\odot_j})^2}}
\end{equation}
for the phonon modes $n,j \in \{\text{TA}, \text{TO1}, \text{LA}, \text{LO1}\}$ that contribute most strongly to the heat transport.

In the ballistic case, there is no interaction between different phonon modes and therefore each mode's behavior is wave-like with unique characteristic frequencies determined by the dot product between the mode's group velocity and grating vector, leading to the complex patterns shown in Figure~4\,(g)-(i) and thus their correlation coefficient is rather low.
The hydrodynamic case, in contrast, requires inter-mode scattering events that are dominantly momentum-conserving.
That results in a collective behavior between the modes and the correlation coefficient is close to unity across the whole matrix.
Lastly, the correlations between the modes in the RTA dataset are displayed, where out-of-equilibrium inter-mode scattering is suppressed and each mode has their own independent relaxation time.
In this case, since the relaxation times of some phonon modes do not differ significantly from one another \cite{ding2018phonon}, the correlations are found to be larger, particularly for the TO and TA modes, than in the ballistic case, but still lower than in the \emph{synchronized} hydrodynamic case.

\section{Conclusions}
Solving the full scattering matrix BTE and subsequently simulating diffuse scattering patterns allows us to determine experimental UEDS signatures for various phonon transport regimes in a pump-probe experiment.
The phonon hydrodynamic regime exhibits dramatic oscillatory signatures in both the transient behavior of Bragg peak intensities and the phonon-diffuse scattering whose frequency is unrelated to the group velocity of acoustic phonon modes.
One can extract momentum and branch resolved transient phonon populations during a second sound oscillation from such diffuse scattering data, demonstrating the novelty of UEDS for studying phonon transport processes.
We show through correlation matrices that this approach can lead to a more complete microscopic understanding of the mechanisms in the hydrodynamic phonon transport regime when compared with previous experimental techniques.
Ultimately, experimental diffraction data will allow us to further test and refine the presented method and apply it to other problems like that of thermal transport in quantum dot structures.
A wide range of materials can be treated within the BTE approach described, as long as their third-order force constants can be computed with reasonable computational resources.

\section{Supplementary Material}
The supplementary material to this article includes a document with the phonon band structure of graphite at a temperature of 100\,K and the deviations in energy stored in the phonon modes for different simulation conditions and timesteps.
Further there is a video that shows the time evolution of the scattering contribution from the TA phonon mode in accordance with the snapshots of Figure~\ref{fig:correlations}\,(a)-(i).

\section{Acknowledgments}
This work was supported by the NSERC Discovery Grants Program under Grant Nos. RGPIN-2021-02957 (S.C.H.) and RGPIN-2019-06001 (B.J.S.).
This work was also supported by the Fonds de Recherche du Québec-Nature et Technologies (FRQNT), the Canada Foundation for Innovation (CFI) and the National Research Council of Canada (NRC) Collaborative R\&D program (Quantum Sensors).
L.K. gratefully acknowledges the support from an FRQNT Merit fellowship.

\section{Data Availability}
The data that support the findings of this study are available from the corresponding author upon reasonable request.

\section{References}
\bibliography{main}

\end{document}